\documentclass[onecolumn]{elsarticle}
\newcommand{\chid}{\chi_\text{disc}}

\newcommand{\chigd}{\chi_\text{5,disc}}

\usepackage{epsfig}
\usepackage{latexsym,amsmath,amssymb,amstext}
\begin{document}
\title{The topological susceptibility in finite temperature QCD and axion cosmology}
\author{Peter Petreczky, Hans-Peter Schadler and Sayantan Sharma}
\address{Physics Department, Brookhaven National Laboratory, Upton, NY 11973, USA}
\begin{abstract}
We study the topological susceptibility in 2+1 flavor QCD above the 
chiral crossover transition temperature using Highly Improved Staggered 
Quark action and several lattice spacings corresponding to temporal 
extent of the lattice, $N_\tau=6,8,10$ and $12$. We observe very 
distinct temperature dependences of the topological susceptibility 
in the ranges above and below $250$ MeV. While for temperatures 
above $250$ MeV, the dependence is found to be consistent 
with dilute instanton gas approximation, at lower temperatures 
the fall-off of topological susceptibility is milder. We discuss the 
consequence of our results for cosmology wherein we estimate the bounds 
on the axion decay constant and the oscillation temperature if indeed the 
QCD axion is a possible dark matter candidate. 
\end{abstract}
\maketitle
\newpage

\section{Introduction}
The visible matter consisting of protons and neutrons derive their masses due to 
strong interactions described by Quantum Chromodynamics (QCD). Advances in non-perturbative 
techniques, mainly Lattice gauge theory has lead to unraveling of the phase diagram of QCD at high 
temperatures and first principles determination of hadron masses and decay constants ( see e.g. 
Refs. \cite{Petreczky:2012rq,Ding:2015ona,Bazavov:2009bb} for recent reviews). One aspect of 
QCD which still lacks a comprehensive understanding is its intricate vacuum structure. 

The QCD vacuum is believed to consist of energy degenerate but topologically distinct minima 
which are separated by potential barriers of height $\Lambda_{QCD}/\alpha_s$, where $\alpha_s$ 
is the coupling constant for strong interactions \cite{Callan:1976je,Jackiw:1976pf}. Each minima 
is characterized by a distinct integer or Chern-Simons number which represent the third homotopy class 
of the mapping between the Euclidean spacetime $\mathcal R^4$ and the gauge manifold of $SU(3)$. Tunneling 
between adjacent minima results in change of Chern-Simons number by unity. Such solutions are called 
instantons. They minimize the classical action density and are topologically protected \cite{Belavin:1975fg}.  
At finite temperature, much richer topological structures are expected to arise. Due to periodicity along 
the Euclidean time-like direction, multi-instanton solutions can exist \cite{Harrington:1978ve} 
which are called the calorons \cite{Kraan:1998pm,Lee:1998bb}. Calorons are characterized 
by a definite topological charge $\pm 1$ as well as the holonomy, which is the the Polyakov loop at spatial
infinity. In presence of a non-trivial holonomy, the caloron solutions in $SU(N_c)$ gauge theory can be 
decomposed into $N_c$ monopoles carrying both electric and magnetic charges, called the dyons 
\cite{Kraan:1998pm, Diakonov:2004jn}. On the other hand at temperatures much higher than $T_c$, 
Debye screening would only allow instantons of radius $\leq 1/T$ to exist and the QCD partition 
function written in terms of non-interacting instantons, is well defined and does not suffer from 
infrared singularities \cite{Gross:1980br}. This is the dilute instanton gas approximation (DIGA).

The outstanding issues are to understand the mechanism of confinement and chiral symmetry breaking in 
terms of these constituents. Some aspects of chiral symmetry breaking can be understood within the 
so called interacting instanton liquid model~\cite{Schafer:1996wv}, but it cannot explain 
confinement. There are hints that dyons could be a promising missing link in our understanding of 
confinement mechanism~\cite{Diakonov:2004jn} and preliminary signals for the existence of dyons have 
been reported \cite{Bornyakov:2013iva} from lattice studies. The other important problem is to 
understand from first-principles whether DIGA can indeed describe QCD vacuum at high 
temperatures. Lattice studies in pure $SU(2)$ and $SU(3)$ gauge theories have observed the 
onset of a dilute instanton gas at $\sim 2~ T_d$, $T_d$ being the deconfinement temperature 
\cite{Edwards:1999zm}. However in QCD, the situation is more richer as index theorem ensures 
that instanton contributions to the partition function are suppressed due to the fermion 
determinant. Presence of light fermions thus may lead to much stronger interactions between 
the instantons or its dyonic constituents. Due to conceptual and computational complexity of 
lattice fermion formulations with exact chiral symmetry, \cite{Kaplan:1992bt,Narayanan:1993ss} 
the QCD vacuum at finite temperature in presence of dynamical fermions have been studied in detail 
only recently \cite{Bazavov:2012qja,Cossu:2013uua,Buchoff:2013nra,Dick:2015twa}. Though some of the 
lattice studies on the low-lying eigenvalue spectrum of the Dirac operator with nearly perfect 
chiral fermions report the onset of dilute instanton gas picture in QCD already near $T_c$ 
\cite{Buchoff:2013nra}, a more recent study on larger lattice volumes and nearly physical quark 
masses, observe onset of the dilute instanton gas scenario at a higher temperature $\sim 1.5~ T_c$~
\cite{Dick:2015twa}. Here $T_c$ denotes the chiral crossover transition temperature, $T_c=154(9)$ MeV 
\cite{Bazavov:2011nk}. It is thus important to study a different observable which is sensitive to 
the topological structures in QCD, to address when can the QCD vacuum be described as a dilute gas 
of instantons and  whether dyons indeed exist near $T_c$. One such observable is the 
topological susceptibility.

Understanding the topological susceptibility, or more generally the structure of QCD vacuum at $T>0$ 
is not only of great conceptual importance but has applications in cosmology as well. One of
the topics that has received renewed interest is whether the axions could be a possible dark matter 
candidate. Strong bounds exist for the magnitude of QCD $\theta$ angle that breaks CP symmetry explicitly and 
measurement of neutron electric-dipole moment has set the value of $\theta<10^{-10}$. One possible way to 
generate such a small value of the QCD $\theta$ angle is through the Peccei-Quinn (PQ) mechanism 
\cite{Peccei:1977hh,Peccei:1977ur}. This involves including a dynamical field $a(x)$ called axion, which has a 
global $U(1)$ symmetry and couples to the gauge fields in the QCD Lagrangian as,
\begin{equation}
 \mathcal{L}_{PQ}=-\frac{1}{4} F_{\mu,\nu} F^{\mu,\nu}-\frac{g^2}{16 \pi^2}\left( \theta+\frac{a(x)}{f}\right)
 F^a_{\mu,\nu}\tilde F^{\mu,\nu,a}+ \frac{1}{2}\partial_\mu a\partial^\mu a~.
\end{equation}
Due to dynamical symmetry breaking of this anomalous $U(1)$ PQ symmetry, the axion field $a(x)$ acquires a vacuum 
expectation value at the global minima given by $\theta+\langle a\rangle/f=0$. The curvature of the axion 
potential or equivalently the free energy, near the global minima is quantified by the topological susceptibility of
QCD, thus generating a mass $m_a$ for the axions. The mass of axions is of central importance~\cite{Vysotsky:1978dc}, 
if indeed the axion is a possible cold dark matter candidate \cite{Turner:1989vc}. At very high temperatures or at 
large scales when inflation sets in, the axion field would have possibly taken random values at causally disconnected 
spacetime coordinates. As the universe expands and cools down the axion field feels the tilt in the potential and 
slowly rolls down towards the global minima. At a temperature $T_{osc}$, where the Hubble scale $H(T_{osc})=m_a(T_{osc})/3$, 
the axion field would start oscillating about its minima. If the amplitude of the oscillations are large, damped only due 
to the Hubble expansion, these largely coherent fields would be a candidate for cold dark matter 
\cite{Preskill:1982cy,Dine:1982ah,Abbott:1982af}. Hence the abundance of axion dark matter is sensitive to the mass of the axion and thus 
to the topological susceptibility in QCD at the scale $T\sim T_{osc}$.  The typical value for $T_{osc} \sim$ GeV and it is 
important to understand the temperature dependence of the topological susceptibility at high enough temperatures.

The topological susceptibility, $\chi_t$ has been studied on the lattice both for cold and hot QCD medium. At $T=0$, 
its continuum extrapolated value is known in quenched QCD \cite{DelDebbio:2004ns,Durr:2006ky} as well with dynamical 
fermions \cite{Bazavov:2010xr}. Motivated from DIGA \cite{Gross:1980br}, the $\chi_t$ can have a power law dependence 
on temperature which is characterized by an exponent $\gamma$, such that $\chi_t\sim T^{-\gamma}$.  Including quantum 
fluctuations to the leading order in $\alpha_s$ over classical instanton action, the exponent $\gamma=11/3 N_c-5+2 N_f/3$ 
for QCD with $N_f$ light quark flavors \cite{Ringwald:1999ze}. In quenched QCD, high statistics study on large volume 
lattices for temperatures upto $2.5 ~T_d$, employing cooling techniques, has reported a $\gamma = 5.64(4)$ 
\cite{Berkowitz:2015aua} which is significantly different from a dilute instanton gas scenario. A more extensive study 
\cite{Borsanyi:2015cka} on a larger temperature range upto $4~ T_d$ using controlled smearing through the Wilson flow, 
has found exponent of $\chi_t$ to be $\gamma=7.1(4)(2)$ in agreement with DIGA. However a prefactor of $\mathcal O(10)$ 
is needed to match the one-loop calculation of $\chi_t$ within DIGA to the lattice data. Extension to QCD with realistic 
quark masses has been discussed \cite{Kitano:2015fla} and very recently performed \cite{Bonati:2015vqz}. It was found in 
Ref.~\cite{Bonati:2015vqz} that the temperature dependence of $\chi_t$ is characterized by $\gamma\sim 2.90(65)$, in 
clear disagreement with DIGA, which predicts $\gamma=8$ for $N_f=3$. This hints to the fact that the $\chi_t$ in QCD with 
realistic quark masses could have a very different temperature dependence than in pure gauge theory.

In order to have a conclusive statement about the temperature dependence of $\chi_t$ in QCD with 
dynamical physical quarks, we approach the problem in a two fold way. We use a very large ensemble of 
QCD configurations on the lattice with Highly Improved staggered Quark (HISQ) discretization and measure 
$\chi_t$ from the winding number of gauge fields after smearing out the ultraviolet fluctuations.
For the staggered quarks, the index theorem is not uniquely defined hence we carefully perform a continuum 
extrapolation of our measured $\chi_t$ for very fine lattice spacings.  We also show the the robustness of 
our findings and indirectly study the validity of the index theorem for staggered quarks, as we approach the 
continuum limit, by comparing the results for $\chi_t$ with those obtained from an observable constructed out 
of fermionic operators, namely the disconnected part of the chiral susceptibility.

We also use our results for $\chi_t$ as a tool to estimate the nature and abundance of the 
topological structures in QCD as a function of temperature.  We show for the first time that 
the effective exponent $\gamma$ governing the fall-off of $\chi_t$ changes around $1.5~T_c$.
The change in the exponent is very robust and persists as we go to finer lattice spacings. 
This may indicate that the nature of topological objects contributing to $\chi_t$ is different 
for $T< 1.5~T_c$ and $T>1.5~T_c$, further supporting the conclusions of Ref. \cite{Dick:2015twa} 
about the onset of diluteness in the instanton ensemble.  Moreover it points to the fact that 
fluctuations in local topology at temperatures close to chiral crossover transition could  
additionally be due to topological objects like the dyons.

The brief outline of the Letter is as follows. In the next section we discuss the 
details of our numerical calculations. Following this are our key results.  The continuum extrapolated result 
for $\chi_t$ is shown and compared with the disconnected chiral susceptibility. The importance of performing 
correct continuum extrapolation to get reliable estimates is discussed in detail. Finally we compare 
our results with that of $\chi_t$ calculated within DIGA considering quantum fluctuations up to two-loop order. 
We show that the dependence of $\chi_t$ in QCD on the renormalization scale is mild and discuss its consequences 
for axion phenomenology. We conclude with an outlook and possible extensions of this work.

\section{Lattice Setup}
We use in this work, the $2+1$ flavor gauge configurations generated by the HotQCD collaboration for determining the 
QCD Equation of State \cite{Bazavov:2014pvz}. These were generated using the HISQ discretization for fermions which 
is known to have the smallest taste symmetry breaking effects among the different known staggered discretizations.
The strange quark mass $m_s$, is physical and the light quark mass is $m_s/20$, which corresponds to a 
Goldstone pion mass of $160~$MeV. The temporal extent of the lattices considered in this work are $N_\tau=6,8,10,12$ 
and the corresponding spatial extents $N_s=4 N_\tau$, such that we are within the thermodynamic limit. For the coarsest 
$N_\tau=6$ lattice, we consider a temperature range between $150$-$800$ MeV and for the finest $N_\tau=12$ lattice, we 
study between $150$-$407$ MeV. We perform gradient flow \cite{Luscher:2010iy} on the gauge ensembles to measure $\chi_t$. 
Typically we have analyzed about $500$ configurations near $T_c$ and to a maximum 
of $13000$ configurations at the higher temperatures and finer lattices. We usually perform 
measurements on gauge configurations separated by 10 RHMC trajectories to avoid autocorrelation 
effects. The number of configurations analyzed for each $\beta=10/g^2$ value are summarized in 
table \ref{tab:table1}.
\begin{center}
\begin{table}[h]
        \begin{tabular}{|l|l|l|l|l|}\hline
           $\beta$ & $N_\tau=6$& $N_\tau=8$  & $N_\tau=10$ & $N_\tau=12$ \\ \hline
        & T (MeV)$\vert$ \# cf      & T (MeV)$\vert$  \# cf     & T (MeV)$\vert$ \# cf      &  T (MeV)$\vert$\# cf  
        \\\hline 
6.608	& ~~266.1~~~$\vert$	500	& ~~199.5~~~$\vert$	500	& ~~159.6~~~$\vert$	500	&	~~~~~~~~~~~--	\\	\hline
6.664	& ~~281.0~~~$\vert$	500	& ~~210.7~~~$\vert$	1000	& ~~168.6~~~$\vert$	500	& ~~149.0~~~$\vert$	320	\\	\hline
6.800	& ~~320.4~~~$\vert$	500	& ~~240.3~~~$\vert$	500	& ~~192.2~~~$\vert$	500	& ~~160.2~~~$\vert$	490	\\	\hline
6.880	&	~~~~~~~~~~--	& ~~259.3~~~$\vert$	500	& ~~207.5~~~$\vert$	970	& ~~172.9~~~$\vert$	500	\\	\hline
6.950	& ~~369.5~~~$\vert$  2970	& ~~277.2~~~$\vert$	500	& ~~221.8~~~$\vert$	500	& ~~184.8~~~$\vert$	500	\\	\hline
7.030	&	~~~~~~~~~~--	& ~~298.9~~~$\vert$ 500	& ~~239.1~~~$\vert$	2790	& ~~199.2~~~$\vert$	2120	\\	\hline
7.150	& ~~445.6~~~$\vert$	2490	& ~~334.2~~~$\vert$	4080	& ~~267.4~~~$\vert$	2650	& ~~222.8~~~$\vert$	3910	\\	\hline
7.280	& ~~502.1~~~$\vert$	3180	& ~~376.6~~~$\vert$	2190	& ~~301.3~~~$\vert$	3830	& ~~251.1~~~$\vert$	5140	\\	\hline
7.373	& ~~546.3~~~$\vert$	1320	& ~~409.7~~~$\vert$	6610	& ~~327.8~~~$\vert$	5550	& ~~273.1~~~$\vert$	13020	\\	\hline
7.596	& ~~666.7~~~$\vert$	1370	& ~~500.0~~~$\vert$	8770	& ~~400.0~~~$\vert$	6770	& ~~333.3~~~$\vert$	10910	\\	\hline
7.825	& ~~814.8~~~$\vert$	1260	& ~~611.1~~~$\vert$	1510	& ~~488.9~~~$\vert$	9050	& ~~407.4~~~$\vert$	6290	\\	\hline
         
        \end{tabular} 
    \caption{The number of gauge configurations (\#cf) analyzed for different $N_\tau$, $T$ and $\beta$ values. }
    \label{tab:table1}
\end{table}
\end{center}
\subsection{Details of our smoothening procedure}
In order to calculate the the topological charge and its fluctuations it
is necessary to get rid of the ultraviolet fluctuations of the gauge fields 
on the lattice. We employ gradient flow to do so. The gradient flow is defined 
by the differential equation of motion of $V_t$, which using same notations as 
in Ref. \cite{Luscher:2010iy}, is 
\begin{equation}\label{eq:grad}
        \partial_t V_t(x,\mu) = -g_0^2(\partial_{x,\mu}S[V_t])V_t(x,\mu) \; ,
\end{equation}
where $g_0$ is the bare gauge coupling and $S[V_t]$ is the Yang-Mills action.
The field variables $V_t(x=(\mathbf{x},\tau),\mu)$ are defined on the four 
dimensional lattice and satisfy the condition $V_t(x,\mu)|_{t=0} = U_{\mu}(x)$,
where $U_{\mu}(x)$ are the usual $SU(3)$ link variables in QCD 
and $t$ is a new coordinate that denotes the flow time which has dimension of $[a^2]$. 
We use the Symanzik flow \cite{Fodor:2014cpa}, i.e. $S[V_t]$ in Eq. \ref{eq:grad} is 
chosen to be the tree-level improved Symanzik gauge action; another possible choice is 
to consider flow with Wilson gauge action \cite{Luscher:2010iy}. For the evolution 
of gauge configurations and calculation of the topological charge we use the publicly 
available MILC code \cite{MILC}.

Since Eq.~(\ref{eq:grad}) has the form of a diffusion equation, the gradient flow results in 
the smearing of the original field $U_{\mu}(x)$ at a length scale $f=\sqrt{8t}$. However, 
unlike the common smearing procedures (see e.g. Refs. \cite{Albanese:1987ds,Hasenfratz:2001hp}), 
which can be applied only iteratively in discrete smearing steps, the amount of smearing in the gradient 
flow can be increased in infinitesimally small steps. This ensures that the length scale over which the 
gauge fields are smeared can be appropriately adjusted as we vary the lattice spacing and/or temperature.
We use a step-size of $0.01$ in lattice units for flow time $t$, which enables us to perform sufficiently 
fine smearing. At $T=0$, the flow time has to be sufficiently large to get rid of ultraviolet noise
but typically has to be smaller than $1/\Lambda_{QCD}^2$. For non-zero temperature, additionally, 
one has to choose the flow time such that $f$ is smaller than $1/T$ 
\cite{Datta:2015bzm,Petreczky:2015yta,Asakawa:2013laa}. In this study we use a maximum of $200$ flow steps
in the calculation of topological charge. This corresponds to the value of $f T= 0.333$ for our finest 
$N_{\tau}=12$ lattices. Even for the coarsest $N_{\tau}=6$ lattices we ensure that $f T< 1$.

\begin{figure}[h]
    \begin{center}
        \includegraphics[width=6cm]{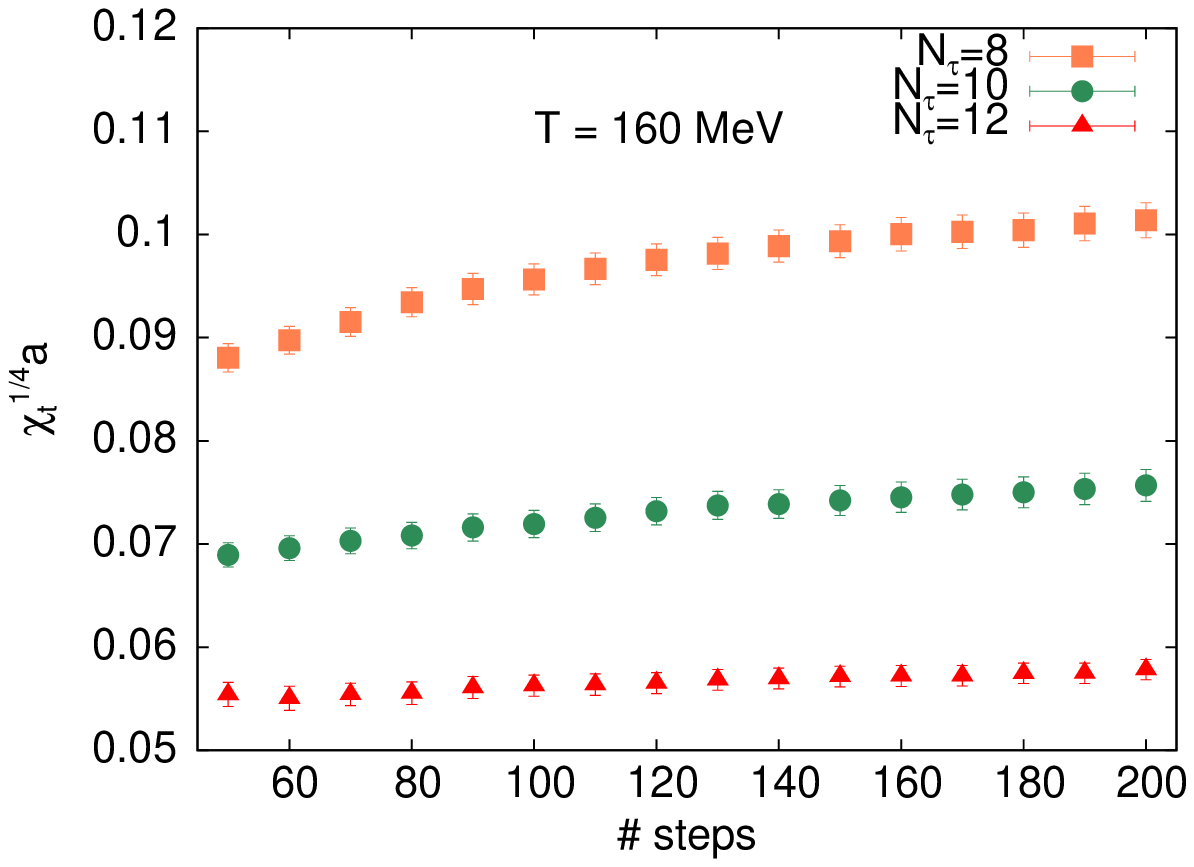}
        \includegraphics[width=6cm]{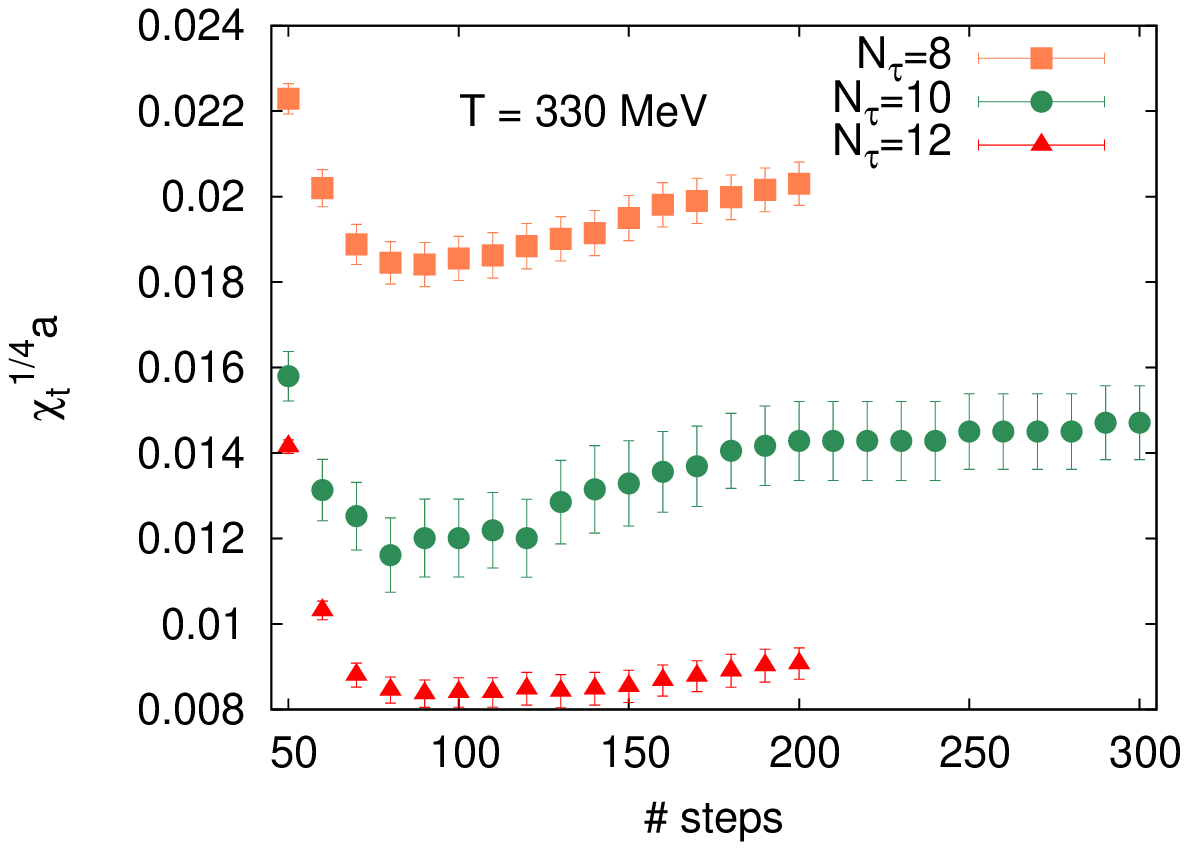}
        \caption{The $N_\tau$ dependence of the topological susceptibility as function of the number of steps 
        of Symanzik flow for two different temperatures near $T_c$ and  $T\sim 330$ MeV. For $N_\tau=10$ and 
        $T=330$ MeV, we show the robustness of the plateau-like region for $\chi_t^{1/4}$ even after $300$ steps 
        of smearing.}
        \label{fig:topvsntvss}
    \end{center}
\end{figure}

At each flow step, the topological charge $Q$ was measured using the clover improved definition 
of $F_{\mu,\nu}\tilde F^{\mu,\nu}$. Fractional values of $Q$ which are within $20\%$ of the next 
higher integer were approximated to that integer. We note that increasing flow time beyond $ft=0.1$, 
the value of $\chi_t$ first increases with increasing flow time and then stabilizes. This is because 
on the smoothened gauge configurations, $\chi_t$ is rounded to the nearest non-zero integer more frequently. 
Since the flow time is never too large i.e. $f T<1$, instantons are not washed due to smearing
 We determined the optimal number of flow steps at each temperature from the plateau of the $\chi_t$. For $T<300$ MeV, 
the plateau in $\chi_t^{1/4}$ was already visible within $50-100$ steps of Symanzik flow. 
At higher temperatures, more than $100$ steps in flow time are required to obtain 
a stable plateau in $\chi_t^{1/4}$.  We compare the onset of the plateau-like behavior 
for $\chi_t^{1/4}$ as a function of the number step sizes for different lattice spacings in 
Fig.~\ref{fig:topvsntvss}. From these data, we infer that $200$ steps of smearing 
is sufficient to give a stable value of the $\chi_t^{1/4}$.  We also checked that for a finer 
$N_\tau=10$ lattice, increasing the number of smearing steps to $300$ did not change the 
value of $\chi_t^{1/4}$. This is shown in the right panel of Fig.~\ref{fig:topvsntvss}. We 
therefore uniformly consider only the values of the $\chi_t$ measured  after performing $200$ 
steps of Symanzik flow for all temperatures and lattice spacings considered in this work.

\begin{figure}[h]
    \begin{center}
        \includegraphics[width=4.2cm,angle=-90]{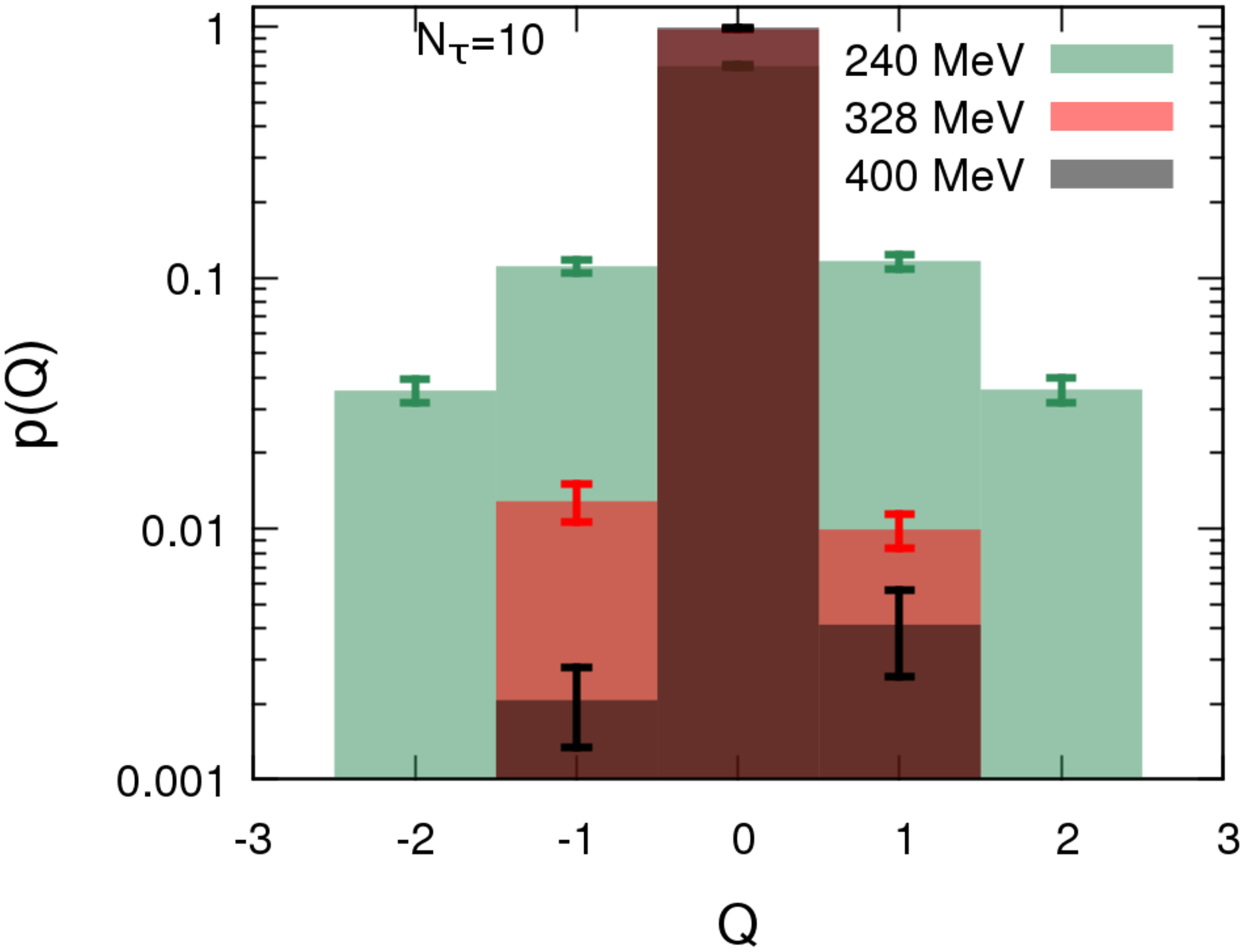}
        \includegraphics[width=4.2cm,angle=-90]{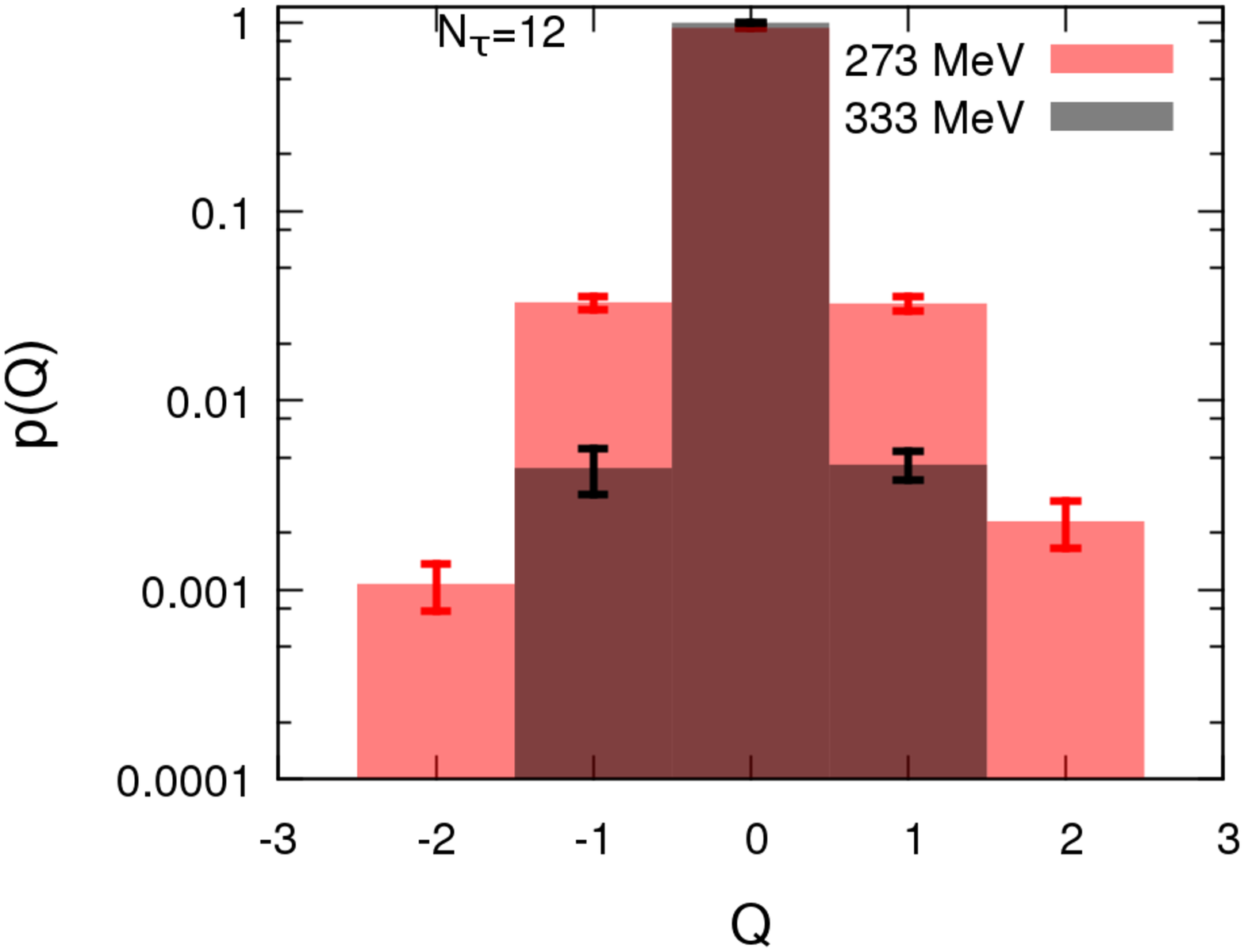}
        \caption{The histograms for the distribution of topological charge at different temperatures and for two different  
        lattice spacings that correspond to $N_\tau=10, 12$ respectively.}
        \label{fig:toptable}
    \end{center}
\end{figure}

In order to ensure that all the topological sectors have been adequately sampled, we measure the topological 
charge distribution for different topological sectors, labeled by $Q$. In Fig.~\ref{fig:toptable} we show the histograms 
for the distribution of $Q$ for our given configurations at different temperatures, upto
 $\sim 2.8~ T_c$. It is evident that different topological sectors have been sampled effectively since the $Q$ 
distribution is a Gaussian peaked at zero at both the temperatures. For $\beta=7.825$, which corresponds to a temperature 
of $\sim 407$ MeV for the $N_\tau=12$ lattice, we still observe tunneling between different topological sectors 
even though the lattice spacing is fine enough, $a=0.041$ fm and lattice volume is large, $V\sim (2.5~\text{fm})^3$. 
However, we do not observe tunneling among the topological sectors for $T>600$ MeV for $N_\tau\geq 8$, even when we 
sample about $\mathcal O(1000)$ gauge configurations.

\section{Results}
\subsection{Topological susceptibility from gradient flow}

\begin{figure}[h]
\includegraphics[width=5.5cm]{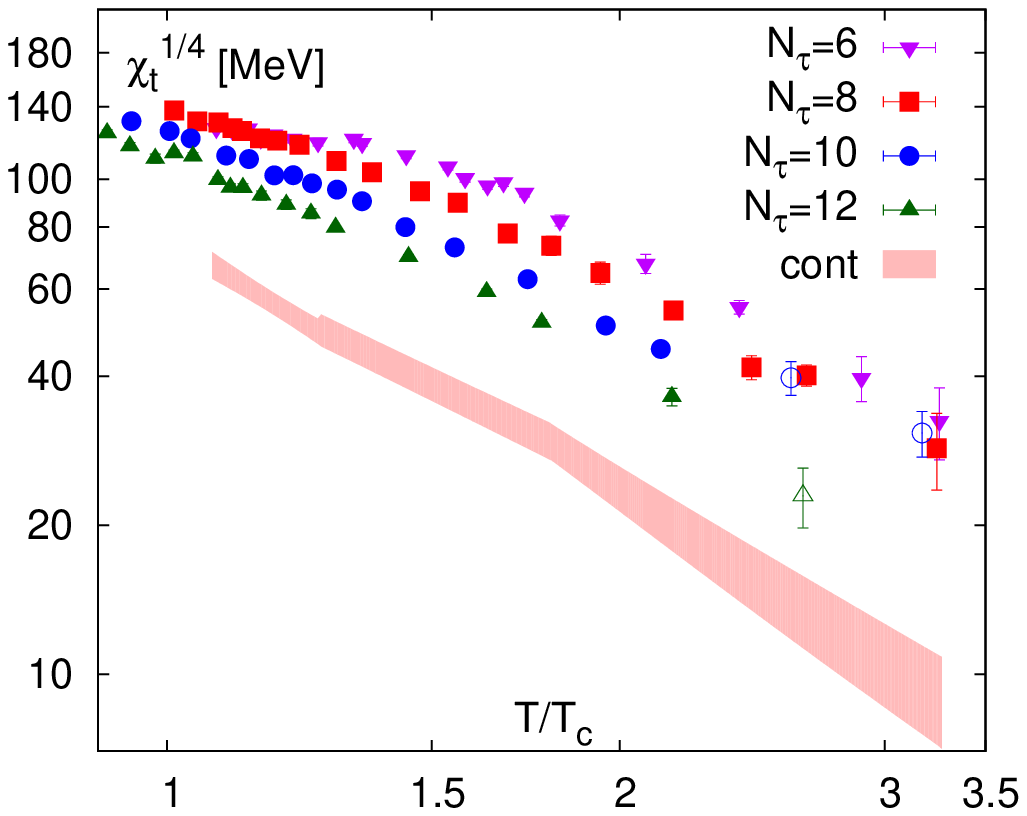}
\includegraphics[width=6.0cm]{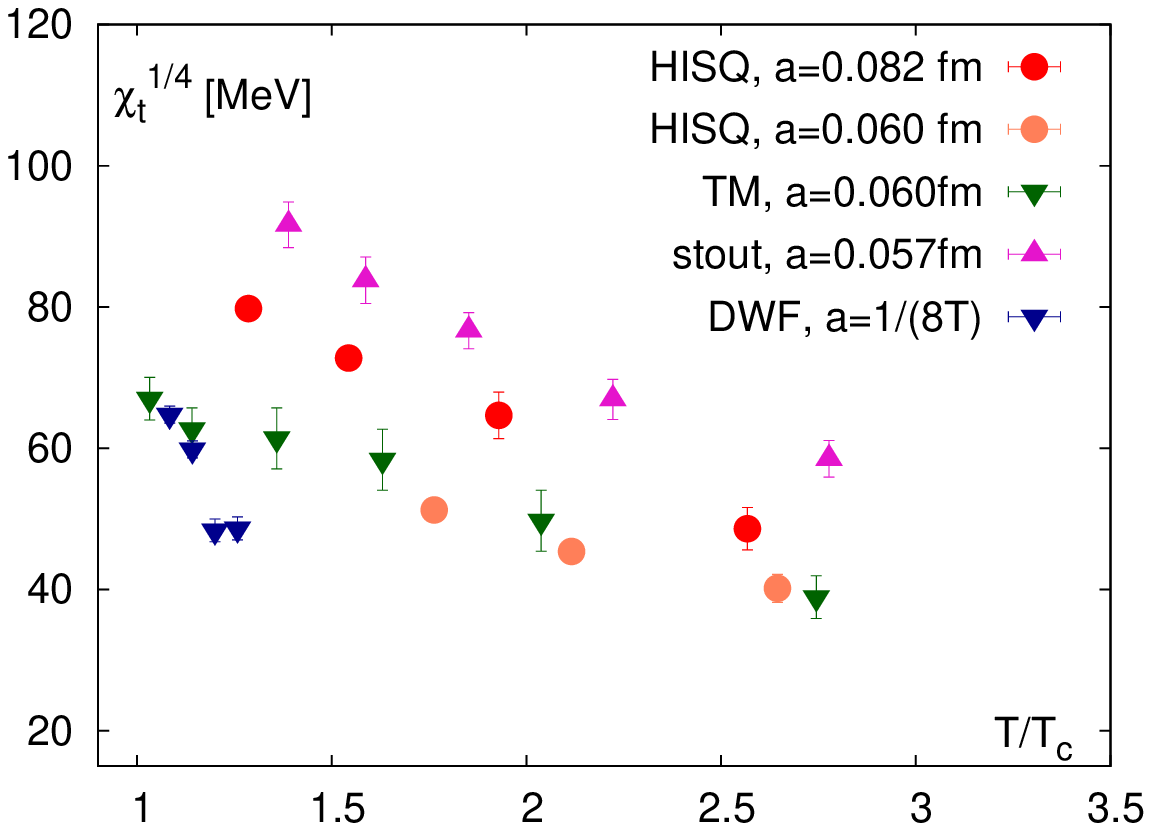}
\caption{The temperature dependence of $\chi_t$ in QCD for HISQ action on lattices with different $N_{\tau}$ (left)
and $\chi_t$ for HISQ action at two lattice spacings compared with recent results using different fermion actions
\cite{Buchoff:2013nra, Bonati:2015vqz,Trunin:2015yda} (right). In the left panel, we also show the continuum
result for $\chi_t$ disucssed in section 4 and open symbols represent the data points that have not been used in
the continuum extrapolation.}
\label{fig:chi_top}
\end{figure}

Our results for the topological susceptibility $\chi_t$ are summarized in Fig. \ref{fig:chi_top}.
In most of our plots we rescale the temperature axis by the chiral crossover transition temperature
$T_c=154$ MeV, unless stated otherwise.
The cut-off effects are reflected in the strong $N_{\tau}$ dependence of the data up to the 
highest temperatures considered. At low temperatures, large cutoff effects in $\chi_t$ can be
understood in terms of breaking of the taste symmetry of staggered fermions and can be quantified 
within the staggered chiral perturbation theory \cite{Bazavov:2009bb,Bazavov:2012xda}. 
However taste breaking effects are expected to be milder at higher temperatures since the lattice 
spacing is finer. Our results indicate that above $T_c$, cutoff effects depend dominantly on $a T=1/N_{\tau}$ 
rather than $ \sim a \Lambda_{QCD}$. In fact, the cutoff effects in $\chi_t$ is much 
stronger than for any other thermodynamic quantity calculated so far with HISQ action 
\cite{Bazavov:2014pvz,Bazavov:2013uja,Ding:2015fca}. It is also evident from our data that $\chi_t$ 
has an interesting temperature dependence. If indeed $\chi_t$ is characterized by a power 
law fall off such that $\chi_t^{1/4}=A T^{-b}$, then a fit to our data restricted to intervals $[165:220]$ 
MeV and $[220:600]$ MeV resulted in $b\sim 0.9$-$1.2$ and $b\sim 1.35$-$1.6$ respectively.
In fact, it turns out that it is impossible to obtain an acceptable fit to the data with a single
exponent $b$ in the entire temperature range.
This led us to model the variation of the exponent by giving it two different values for temperatures 
$T > 1.5 T_c$ and $T < 1.5 T_c$. This is also taken into account when performing continuum extrapolation. 
We note that at leading order, the exponent calculated within DIGA for $N_f=3$ corresponds to $b=\gamma/4=2$. 
Higher order corrections would effectively reduce the value of $b$.

Let us compare our results with those obtained in QCD using different fermion discretizations 
\cite{Buchoff:2013nra, Bonati:2015vqz,Trunin:2015yda}. We do not expect the values of $\chi_t$ for 
different choice of fermion discretizations to agree with each other at non-vanishing lattice spacings. 
However such comparisons provide insights on the nature and severity of the discretization effects. 
Another source of this difference could be due to the fact that all earlier calculations have been 
performed for different values of quark masses. For a fair comparison with the earlier results, we 
scale $\chi_t $ with $m_l^2 \sim m_{\pi}^4$.  
In Fig. \ref{fig:chi_top} we show our results for two lattice spacings $a=0.082$ fm and $a=0.06$ fm
and compare them with other recent results for $\chi_t$. Our results for $a=0.06$ fm agree well
with the ones obtained using twisted mass Wilson fermions for pion mass $370$ MeV and very similar
lattice spacing once rescaling with the pion mass is performed \cite{Trunin:2015yda}.
Comparing our results with the latest calculations performed using
staggered fermions with the so-called stout action \cite{Bonati:2015vqz} and physical pion mass, we find that 
our results are $\sim 70\%$ of the values obtained in Ref. \cite{Bonati:2015vqz} at similar lattice spacings. 
This could be due to the fact that discretization errors for the HISQ action considered in our work are smaller 
than for the stout action. Domain wall fermions give the smallest value of $\chi_t$ even
on relatively coarse lattice, $a=1/(8 T)$, which is not surprising
since the chiral symmetry in the corresponding calculation is almost exact.

\subsection{Disconnected chiral condensate and topological susceptibility}
Low-lying eigenvalues of the QCD Dirac operator are known to be sensitive to 
topological objects of the underlying gauge configurations. Calculating
fermionic observables which are sensitive to these low-lying eigenvalues 
provides an alternative way to study the topological fluctuations in QCD.
One such observable is the disconnected chiral susceptibility. QCD action with 
two light quark flavors has a $SU_A(2)\times SU_V(2)\times U_B(1) \times U_A(1) $
symmetry which is effectively broken to $SU_V(2)\times U_B(1)$ at $T_c$. 
The effective restoration of these symmetries can be studied through the 
volume integrated meson correlation functions or susceptibilities,  
\begin{eqnarray}
\label{eq:chi_sigma_def}
\chi_\sigma &=& \frac{1}{2}\int d^4 x \left\langle \sigma(x) \sigma(0) \right\rangle  ~,~
\chi_\delta   = \frac{1}{2}\int d^4 x \left\langle \delta^i(x) \delta^i(0) \right\rangle  \\
\chi_\eta     & =& \frac{1}{2}\int d^4 x \left\langle \eta(x)     \eta(0) \right\rangle ~,~      
\chi_\pi        = \frac{1}{2}\int d^4 x \left\langle \pi^i(x)      \pi^i(0) \right\rangle.
\end{eqnarray}
The above quantities can be expressed in terms of quark fields which in the one-flavor 
normalization \cite{Buchoff:2013nra} are given as, $  \sigma = \overline{\psi}_l \psi_l~,
~\delta^i = \overline{\psi}_l \tau^i \psi_l ~,~ \eta = i\overline{\psi}_l \gamma^5 \psi_l~,~
\pi^i = i\overline{\psi}_l \tau^i \gamma^5 \psi_l$. The susceptibilities in the different 
quantum number channels are related as, 
\begin{equation}
\label{eq:chid}
  \chi_\sigma = \chi_\delta + 2 \chid ~,~ 
  \chi_\eta = \chi_\pi - 2 \chigd~,  
  \end{equation}
where the $\chi_{\pi,\delta}$ only contain the connected part of the correlators. When 
chiral symmetry is effectively restored in QCD, $\chi_\sigma  =  \chi_\pi $ and
$\chi_\eta =  \chi_\delta$. This implies that the difference in the integrated 
correlator $\chi_\pi - \chi_\delta = 2\chid$. Effective restoration of $U_A(1)$ would 
imply  $\chi_\pi = \chi_\delta$, which results in $\chid=0$. Therefore $\chid$ can be considered 
as the measure for $U_A(1)$ breaking. Expressing the topological charge as the trace of the 
Dirac fermion matrix for light quarks \cite{Buchoff:2013nra}, 
$ Q=m_l~ {\rm Tr} ( \gamma_5 M_l^{-1} )$, its fluctuation normalized by four-volume is simply 
the topological susceptibility defined as,
\begin{equation}
\chi_t = \frac{m_l^2}{N_\sigma^3 N_\tau a^4}  \left\langle\left(\textrm{Tr}M_l^{-1}\gamma^5\right)^2
\right\rangle= (m_l)^2 \chigd  ~,
\label{eq:qsusc}
\end{equation}

\begin{figure}
\includegraphics[width=6cm]{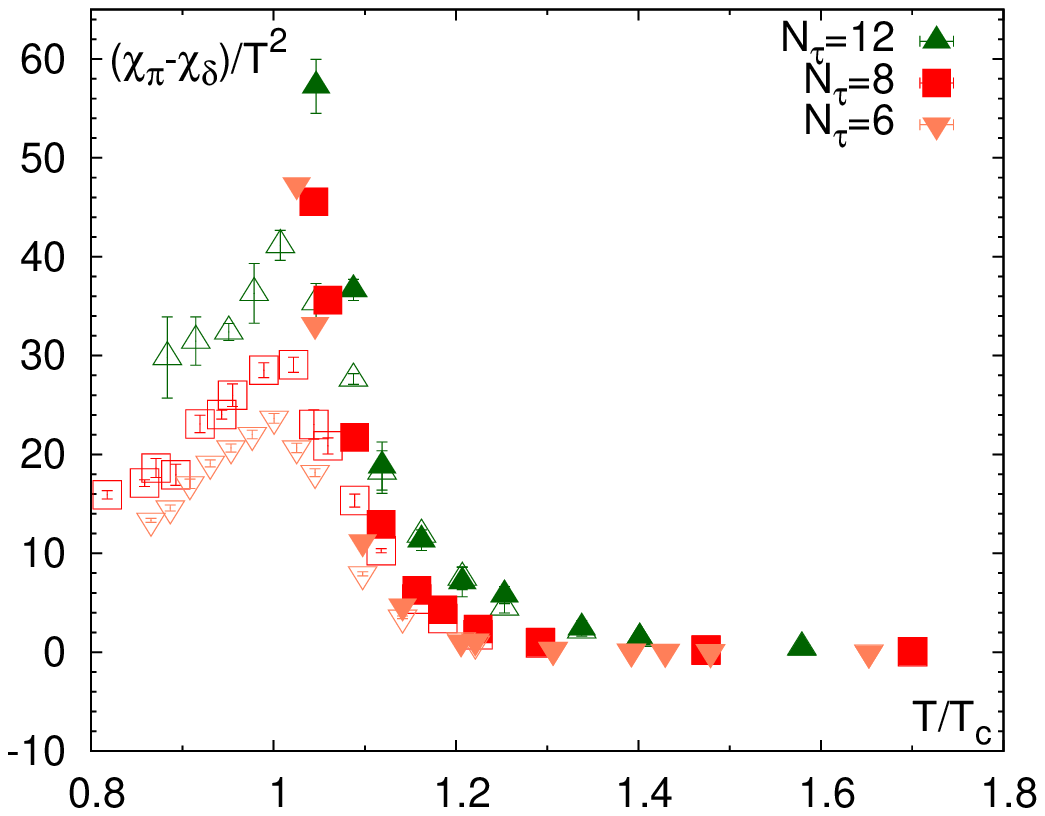}
\includegraphics[width=6cm]{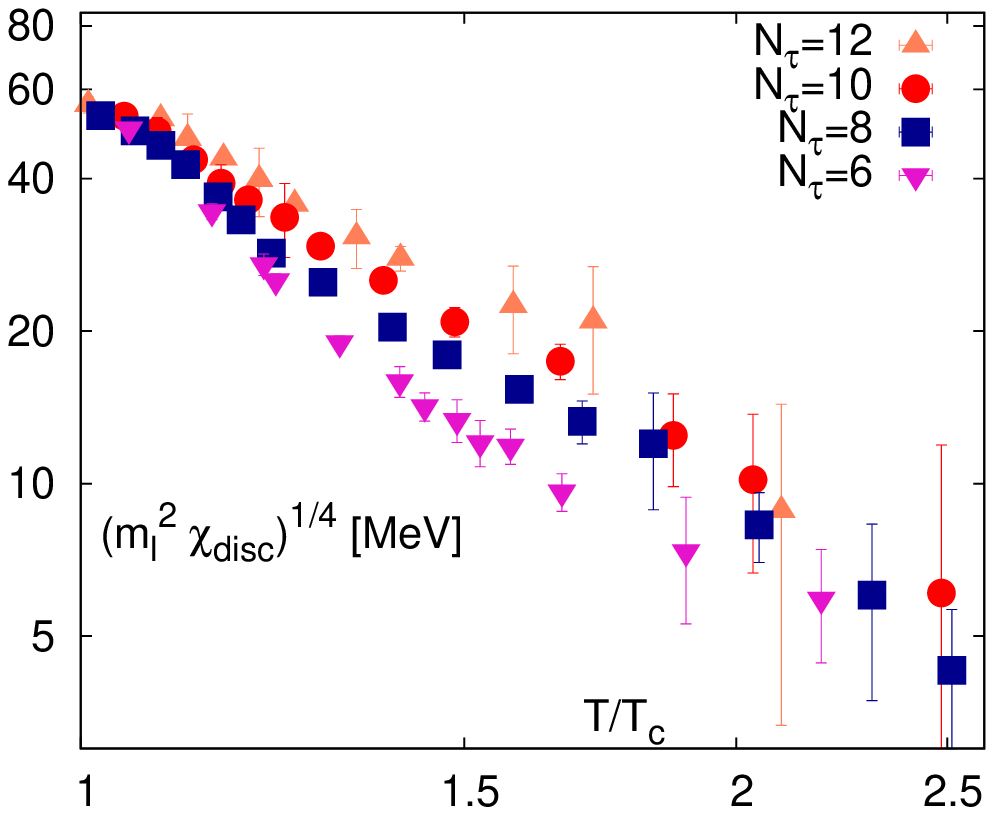}
\caption{The quantities $(\chi_\pi-\chi_\delta)/T^2$ (left) and $(m_l^2 \chid)^{1/4}$ (right) measured on 
the lattice as function of $T/T_c$ for various $N_\tau$. Here the chiral crossover transition temperature 
$T_c=154$ MeV. The open symbols in the left panel correspond to the values of $\chid$.}
\label{fig:diff}
\end{figure}
 
In the chiral symmetry restored phase, $\chi_t=(m_l)^2 \chid $ since $\chigd=\chid$. In  other 
words $\chid$ is a proxy for $\chi_t$. The above considerations are valid on the lattice only 
near the continuum limit.
In order to check how well the corresponding equalities are satisfied at finite lattice spacing, 
we first consider the quantity $\chi_\pi - \chi_\delta$ normalized by $T^2$.
$\chi_\pi$ was estimated from the Chiral Ward identity $\chi_\pi=\langle \bar \psi \psi \rangle_l/m_l$, 
and $\chi_\delta$ from the connected part of meson correlators $\chi_{conn}$ 
from Ref. \cite{Bazavov:2011nk} (Tables X-XII) and converting the numerical
results to single flavor normalization. The results for $\chi_\pi - \chi_\delta$ for our 
QCD ensembles with HISQ discretization are shown in Fig. \ref{fig:diff}. It is evident from this plot 
that for $T \gtrsim 1.2 T_c$ the quantity $\chi_\pi-\chi_\delta=\chid$ for all different lattice 
spacings considered therefore  $\chi_t$ can be independently estimated from the disconnected 2-point 
chiral susceptibility $\chid$. This estimate of $\chi_t$ is shown in Fig. \ref{fig:diff}. We used the combined
data on $\chid$ from Refs. \cite{Bazavov:2011nk} and \cite{Bazavov:2014pvz}, since the later has
high statistics results for a larger temperature range. A closer inspection of the 
lattice data on $\chid$ in Ref. \cite{Bazavov:2014pvz} reveals that the magnitude of the errors on 
each data point increases with increasing temperatures but at some higher temperature, drops abruptly. 
We interpret this fact to the freezing  of topological charge at high temperatures which results in 
unreliable estimates for $\chid$. This is consistent with our measurement of $\chi_t$ from Symanzik flow, 
where we indeed observe a signal for $\chi_t$ beyond $T=500$ MeV for $N_{\tau}=6, 8$ lattices and beyond 
$T>400$ MeV for $N_\tau=10, 12$ lattices but with larger systematic and statistical uncertainties. 
Therefore, the corresponding data points (shown as open symbols in left panel of Fig. 3) are not included 
in the continuum extrapolation discussed in the following section. 
The significant $N_\tau$ dependence of the results for $\chid$ evident from Fig. \ref{fig:diff} 
further validates our observation in the previous section that the dominant cutoff-effects for $\chi_t$ are 
controlled by $a T=1/N_\tau$ rather than the value of the lattice spacing in some physical units like 
$\Lambda_{QCD}$. 
On finite lattices, the values of $m_l^2 \chid$ are considerably smaller than the values of $\chi_t$ defined 
through gauge observables. 
Furthermore, the approach to the continuum for  $m_l^2 \chid$ is opposite to $\chi_t$, determined from the 
fluctuations of winding number of the gauge fields. This significant difference between the values of $\chi_t$ 
measured in terms of gluonic and fermionic observables has been observed earlier \cite{Buchoff:2013nra} using Domain 
wall fermions, which has nearly perfect chiral symmetry on a finite lattice. Therefore this difference is 
unlikely due to taste-breaking effects inherent in staggered quark discretization. In the next section we 
show a reliable method for the continuum extrapolation of these observables even with such large lattice 
artifacts.

\section{Continuum Extrapolation of topological susceptibility}

In this section we discuss in detail our procedure for performing continuum extrapolation of 
$\chi_t^{1/4}$. As noted earlier, the exponent controlling the power law fall-off of $\chi_t^{1/4}$ 
on temperature depends on the chosen temperature interval. We performed simultaneous fits to the 
lattice data obtained for different $N_{\tau}$ with the following ansatz,
\begin{equation}
\chi_t^{1/4}(T,N_{\tau})=(a_0+a_2/N_{\tau}^2+a_4/N_{\tau}^4) \cdot 
(T_c/T)^{b+b_2/N_{\tau}^2+b_4/N_{\tau}^4+b_6/N_{\tau}^6},
\label{fit}
\end{equation}
As mentioned in the previous section, data at high temperatures, where statistical errors
may not have been estimated correctly are excluded from the fit.

\begin{figure}[h]
\includegraphics[width=4.5cm, angle=-90]{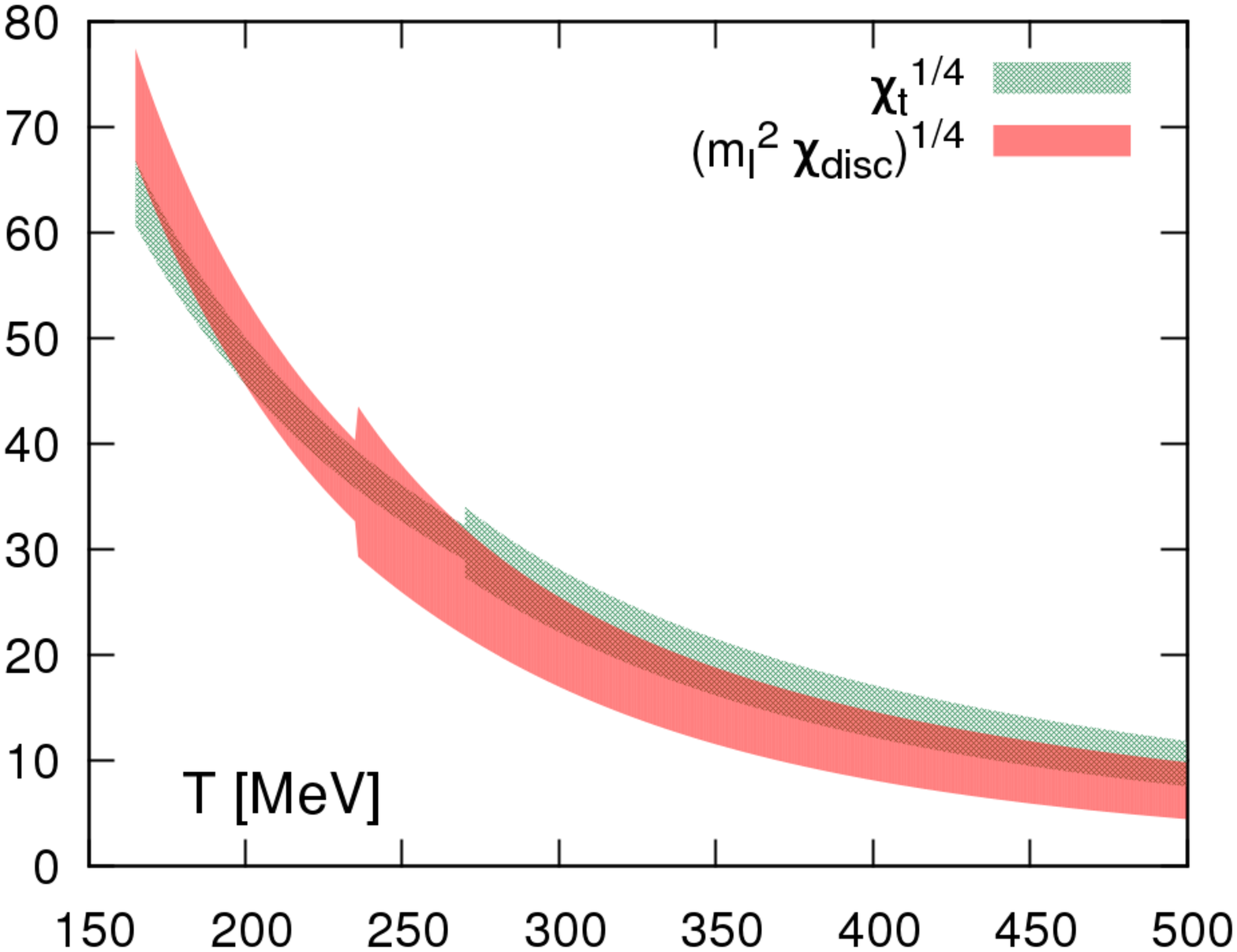}
\includegraphics[width=4.5cm, angle=-90]{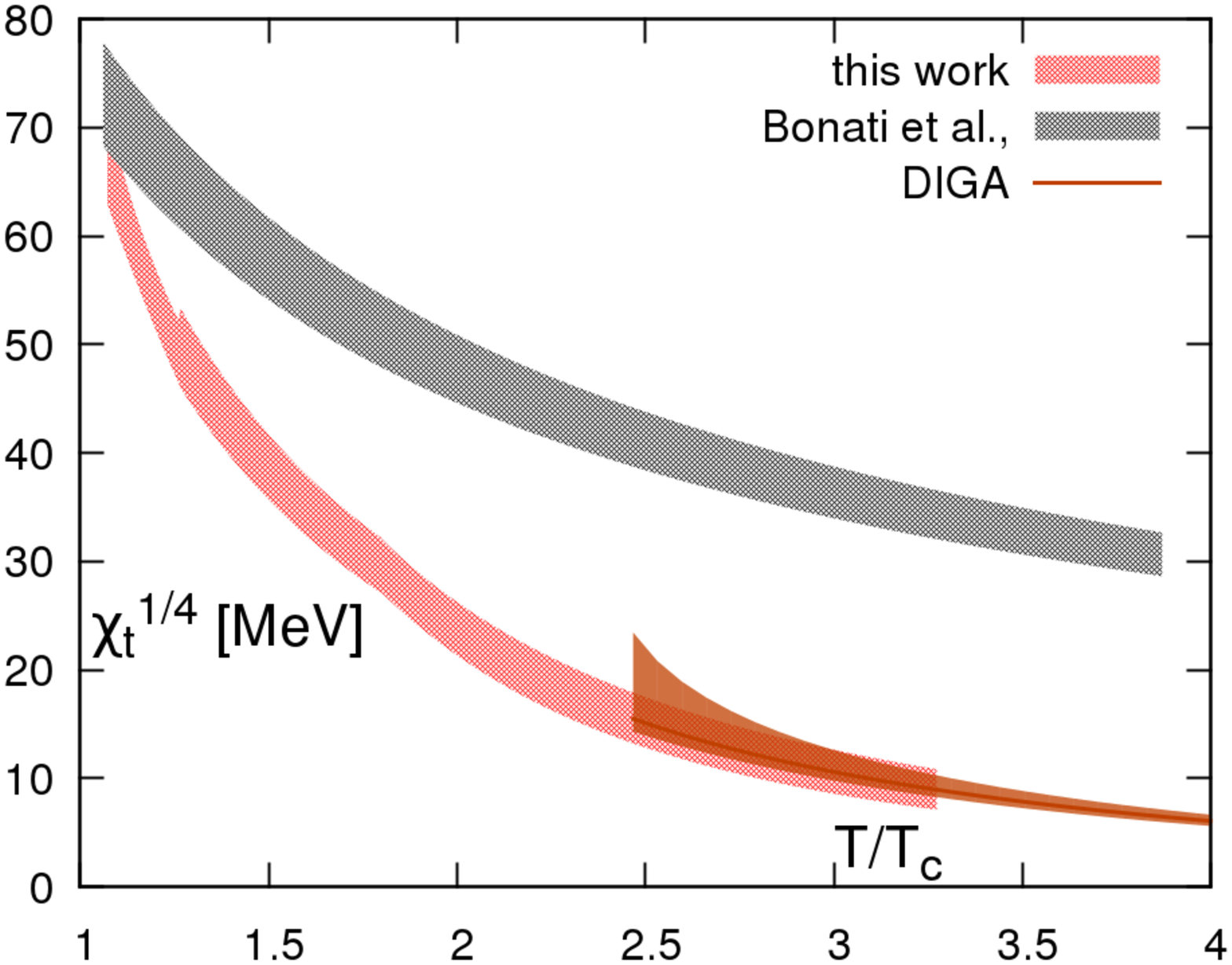}
\caption{Continuum extrapolated results for $\chi_t^{1/4}$ measured from gluonic definition of 
topological charge and $(m_l^2 \chi_{disc})^{1/4}$ fitted separately (left) and the joint continuum 
extrapolation of $\chi_t^{1/4}$ and $(m_l^2 \chi_{disc})^{1/4}$ (right).
In the right panel, our results are compared with the continuum extrapolated results obtained
in Ref. \cite{Bonati:2015vqz}. The solid orange line corresponds to a partial two-loop DIGA calculation 
with $\mu=\pi T$, $K=1.9$ and $\alpha_s(\mu=1.5~{\rm GeV})=0.336$, while the band is obtained  
from the variation of $\alpha_s$ by $1\sigma$ around this central value as well as variation of 
the scale $\mu$ by a factor two (see text).}
\label{fig:cont}
\end{figure}

One of the important issues that needs to be clarified is whether two different estimates of $\chi_t$ 
obtained through the gluonic operator definition for $Q$ and through $m_l^2 \chi_{disc}$ respectively, 
agree in the continuum limit.
We first performed fits to our numerical results of $\chi_t^{1/4}$ presented in section 3.1 according to Eq. \ref{fit}.
We considered two independent fits for the two temperature intervals $165~{\rm MeV} \le T < 270~{\rm MeV}$
and $240~{\rm MeV} \le T \le 504~{\rm MeV}$, respectively. We set $b_4=b_6=0$ and to ensure to be within 
the scaling regime, we omitted the $N_{\tau}=6$ data. 
This did not change the continuum value of $\chi_t$ within errors. Furthermore, when we performed fits setting 
$a_4=0$, the continuum extrapolated value was $15\%-25\%$ larger. This implies that the term proportional to $a_4$ 
is important but subdominant relative to the leading term proportional to $a_2$.
In the first temperature region we obtain $b=1.496(73)$, while the corresponding value from fit to the higher temperature 
interval gave $b=1.85(15)$. 
We found that continuum extrapolations with Eq. \ref{fit} in the entire temperature interval is not possible.
Thus, the temperature dependence of $\chi_t^{1/4}$ is clearly different for these two temperature intervals with the 
exponent for $T>250$ MeV being consistent with the prediction within DIGA. We smoothly interpolate between the 
two different fits at an intermediate $T=270$ MeV, the resulting continuum estimate is shown in Fig. \ref{fig:cont}.

A similar analysis was performed for $(m_l^2 \chi_{disc})^{1/4}$. In this case the low and high temperature 
regions were defined as $165~{\rm MeV} \le T \le 240~{\rm MeV}$ and $240~{\rm MeV} < T \le 504~{\rm MeV}$, 
respectively. From the fits we get $b=1.96(22)$ in the low temperature region, while in the high temperature 
region, the resulting $b=2.22(27)$.  The two fits were matched at $T=235$ MeV to obtain a continuum estimate, shown 
as a red band in Fig. \ref{fig:cont}. It is clearly evident that the continuum estimates for $\chi_t^{1/4}$ from a
gluonic observable and that for $(m_l^2 \chid)^{1/4}$ coincide.  This is highly non-trivial and makes us confident 
that the continuum extrapolation is reliable even though the cutoff effects are important.
A joint fit was also performed according to Eq. \ref{fit}, 
allowing the parameters $a_2,~a_4,~b_2,~b_4$ and $b_6$ to be different for $(m_l^2 \chid)^{1/4}$ and $\chi_t^{1/4}$ 
since the cut-off effects in these quantities are clearly different. We considered three different ranges
$165~{\rm MeV} \le T \le 210~{\rm MeV}$, $195~{\rm MeV} \le T \le 300~{\rm MeV}$ and $230~{\rm MeV} 
\le T \le 504~{\rm MeV}$. We checked different fit ans\"atze, setting some of the parameters $a_i$ 
and $b_i$ to zero and also including or excluding the $N_{\tau}=6$ 
data. All such trials resulted in $\chi^2$ per degree of freedom for the fitting procedure to be about 
one or less. Therefore we simply averaged over all such fit results in each temperature region and used the spread 
in the fit to estimate the final error on our continuum extrapolation. We also checked that the error 
estimated this way is about the same as the statistical error of the representative fits. We matched 
the fit results for the three different regions at $T=194$ MeV and $T=277$ MeV respectively to obtain our 
final continuum estimate for $\chi_t$, shown in right panel of Fig. \ref{fig:cont}. 

Comparing our continuum fit for $\chi_t$ with the continuum extrapolated results from Ref. \cite{Bonati:2015vqz}, 
we find that the exponent $b$ from our fit is about factor three larger than the corresponding $b$ reported there. 
The possible reason for this discrepancy could be due to the fact that $\chi_t$ has been calculated only with 
$N_{\tau}=6, 8$ lattices in Ref. \cite{Bonati:2015vqz} at high temperatures. As discussed earlier, the cutoff 
effects depend on $a T=1/N_{\tau}$ therefore, lattices with $N_{\tau}\geq8$ are needed for reliable continuum 
extrapolation.
To verify this assertion we performed fits to $\chi_t^{1/4}$ data calculated for specific choices of lattice 
spacings, $a=0.0600,~0.0737$ and $0.0825$ fm which are close to the values 
$a=0.0572,~0.0707$ and $0.0824$ fm measured in the earlier work. Using identical fit ansatz 
$(\tilde A_0+a^2 \tilde A_2)(T_c/T)^b$ for $\chi_t^{1/4}$, we get $b=0.674(42)$ for $T>1.3T_c$,  
comparable to $b=0.671(40)$ for $T>1.2T_c$ reported in Ref. \cite{Bonati:2015vqz}. 
We thus conclude that continuum extrapolation procedure for $\chi_t$ is not reliable unless data 
from finer lattice spacings are used.

\section{Instanton Gas and Axion Phenomenology}
At high temperatures, the maximum size of instantons is limited due to the Debye screening 
therefore, the fluctuations of the topological charge can be described within DIGA \cite{Gross:1980br}. 
In this section, we calculate $\chi_t$ in QCD within DIGA following the general outline in 
Refs. \cite{Ringwald:1999ze,Borsanyi:2015cka}. The $\chi_t$ for QCD with $N_f$ quark flavors 
can be written in terms of the density of instantons $D(\rho)$ of size $\rho$ as 
\cite{Ringwald:1999ze,Borsanyi:2015cka},
\begin{equation}
\label{eqn:chitdigm}
\chi_t(T)=2 \int_0^{\infty} d \rho ~D(\rho)~G(\rho \pi T),
\end{equation}
where we use the 2-loop renormalization group (RG) invariant form \cite{Morris:1984zi} of the 
instanton density in ${\overline{MS}}$ scheme
\begin{equation}
D(\rho)=\frac{d_{\overline{MS}}}{\rho^5} \left(\frac{2\pi}{\alpha_s(\mu)}\right)^6 
e^{-\frac{2 \pi}{\alpha_s(\mu)}}
(\rho \mu)^{\beta_0+(\beta_1-12 \beta_0+8 N_f)\frac{\alpha_s}{4 \pi}}
\prod_{i=1}^{N_f} (\rho m_i)~.
\label{eqn:diga}
\end{equation}
Here $G(x)$ is the cut-off function that includes the effects of Debye screening in a 
thermal medium \cite{Gross:1980br} and $\beta_0$ and $\beta_1$ are 
the one and two-loop coefficients of the QCD beta function, respectively.
The constant $d_{\overline{MS}}$ has been calculated in Refs. 
\cite{Hasenfratz:1981tw,Luscher:1981zf,tHooft:1986nc}. At temperatures $T>T_c$, 
we expect the strange quark to contribute to $\chi_t$ so we henceforth consider 
$N_f=3$ in our calculations. At one-loop level, $\chi_t$ calculated from Eq. 
(\ref{eqn:chitdigm}) varies as $\sim T^{-8}$ in QCD with three quark flavors. With the 
two-loop corrections included, the temperature dependence becomes more complicated. 
To evaluate $\chi_t(T)$ according to Eqs. (\ref{eqn:chitdigm},\ref{eqn:diga}) one needs to 
specify $\alpha_s(\mu)$ and the strange quark mass $m_s(\mu)$ in ${\overline{MS}}$ 
scheme as function of the renormalization scale $\mu$. For the running of coupling 
constant, we set the starting value $\alpha_s(\mu=1.5{\rm GeV})=0.336(+12)(-0.008)$ 
obtained from lattice calculations of the static quark anti-quark potential \cite{Bazavov:2014soa}. 
This value is compatible with the earlier determination of $\alpha_s$ \cite{Bazavov:2012ka} 
but has smaller errors. We fix the strange quark mass $m_s(\mu=2{\rm GeV})=93.6$ MeV, as 
determined in Ref. \cite{Chakraborty:2014aca}. Since we are interested in comparing to our 
lattice results, the value of light quark mass was set to $m_l=m_s/20$. For the scale 
dependence of the  $\alpha_s(\mu)$ and quark masses we use  four-loop RG equations as 
implemented in the RunDeC Mathematica package \cite{Chetyrkin:2000yt}.

The central value of the renormalization scale is chosen to be $\mu=\pi T$, motivated from the fact 
that the instanton size is always re-scaled with $\pi T$ in the screening functions $G(\rho \pi T)$.  
To match the partial two-loop result for $\chi_t$ that we calculated within DIGA for $\mu=\pi T$ to 
our lattice results, we have to scale it by a factor $K=1.90(35)$. This factor is obtained by requiring 
that the calculations within DIGA and the continuum extrapolated lattice results agree at $T=450$ MeV.
The value of the $K$-factor in QCD is similar to that obtained in Ref. \cite{Borsanyi:2015cka} for $SU(3)$ 
gauge theory. The comparison between the scaled DIGA and lattice results can be seen from 
Fig. \ref{fig:cont}, where the uncertainty band in the DIGA prediction was estimated by varying the 
renormalization scale $\mu$ by a factor of two and the input value of $\alpha_s(1.5 ~{\rm GeV})$ 
by one sigma.  The strong temperature dependence of the upper bound is due to the fact that the  
lower value of scale $\mu$ becomes quite small when temperatures are decreased, resulting in a  
larger value of $\alpha_s$. 
We note that while the two-loop corrections to the instanton density reduces the uncertainties due 
to the choice of renormalization scale, the difference between the one and two-loop results is not large, 
e.g. for $T=450$ MeV it is only about $1\%$. One may thus wonder about the origin of the large $K$-factor 
needed to match the DIGA estimates with lattice results since the uncertainties due to higher loop 
effects in the instanton density $D(\rho)$ are small. We believe the reason for a large value of 
$K$-factor is inherent in the physics of the high temperature QCD plasma. It is well known that the 
naive perturbative series for thermodynamic quantities is not convergent \cite{Arnold:1994eb,
Bazavov:2016uvm,Berwein:2015ayt}. The situation is particularly worse for the 
Debye mass, where corrections to the leading order result could be as large as the leading 
order result itself \cite{Kajantie:1997pd,Karsch:1998tx,Laine:1997nq} even up to very high
temperatures. Therefore, the calculation of the instanton
density in Ref. \cite{Gross:1980br} should be extended to higher orders within
DIGA for a better comparison with lattice results. This is clearly a very difficult task.
For the Debye mass, higher order corrections do not modify significantly
its temperature dependence, i.e. they appear as a constant multiplicative factor
to the leading order Debye mass. 
Therefore, including higher order effects within DIGA by a $K$-factor appears to be
reasonable and appears to work well in the temperature range probed by the lattice 
calculations.

Let us finally mention that at temperatures of about $1$ GeV, the effects of the charm quarks need 
to be taken into consideration when calculating $D(\rho)$. It is not clear at present, how to 
consistently treat the charm quarks within DIGA since they are neither light nor heavy compared to 
this temperature. Treating charm quarks as light degrees of freedom, the value of $\chi_t$ obtained 
within DIGA increases by $\sim 11 \%$ at $T=1$ GeV compared to the $N_f=3$ result. This is clearly 
a modest effect compared to the large loop effects encoded within the $K$-factor.

\begin{figure}
\begin{center}
\includegraphics[width=5.5cm, angle=-90]{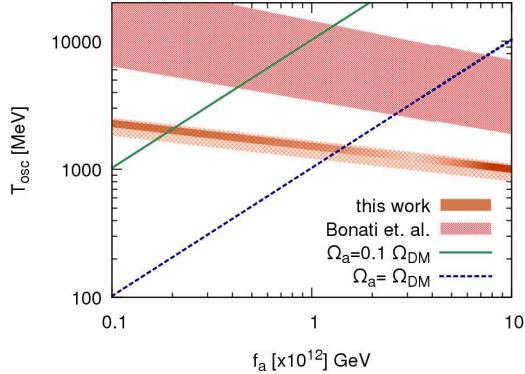}
\caption{ The allowed values of $f_a$ and $T_{osc}$, if indeed the present day dark matter abundance 
is due to the QCD axions. The input from QCD for determining the allowed regions is the magnitude of 
$\chi_t$ at the time of slow roll. The dark-orange band correspond to estimates within DIGA with $K=1.90(35)$ 
that is favored by our lattice data for $\chi_t$ whereas the red band is obtained using the data from  Ref. 
\cite{Bonati:2015vqz}. The light-orange band corresponds to estimates of $\chi_t$ using DIGA with a $K$-factor between 
$1$-$2.5$. The blue (dashed) and green lines denote the upper bounds for $f_a$ for different values of $T_{osc}$, if axions constitute 
the total or $10\%$ respectively, of the present day dark matter abundance (for details see text).}
\label{fig:axiondm}
\end{center}
\end{figure}

We can now predict a bound on the axion decay constant $f_a$, if indeed the cold dark matter abundance
$\Omega_{\text{DM}} h^2$ in the present day universe is due to QCD axions i.e., $\Omega_a h^2\leq 
\Omega_{\text{DM}} h^2$. Assuming that the PQ symmetry breaking and the slow roll of the axion field happened 
after inflation and choosing the misalignment angle $\theta=\langle a \rangle/f_a$, at 
PQ scale to be averaged over different regions of the universe such that $\theta^2=\pi^2/3$, 
we estimate lines of constant $T_{osc}/f_a$ if axions constitute a fraction $r= \Omega_a /\Omega_{\text{DM}}$ 
of the present dark matter abundance.
The lines of constant $T_{osc}/f_a$ for $r=0.1,1$ are denoted by the solid and the dashed lines in Fig. 
\ref{fig:axiondm} respectively, considering the latest PLANCK data~\cite{Ade:2013zuv} for 
$\Omega_{\text{DM}} h^2=0.1199(27)$. For estimating these lines, we assumed that the temperatures are high 
enough for the charm quarks and $\tau$ leptons to be still relativistic, whereas the bottom quarks are 
not considered. Clearly for the values of $T_{osc}$ beyond the bottom quark mass threshold, one has to 
consider them as relativistic degrees of freedom as well, which however will shift the current estimates 
of the lines of constant $T_{osc}/f_a$ by only a few percent. 
From the condition of slow roll $m_a(T_{osc})=3 H(T_{osc})$ and using the relation $m_a^2=\chi_t/f_a^2$, 
where $\chi_t$ is an input from QCD calculations, we can again constrain the allowed regions in the 
$T_{osc}-f_a$ plane. The dark-orange band in Fig. \ref{fig:axiondm} shows the allowed values for 
$f_a$ and $T_{osc}$ within DIGA with a $K$-factor of $1.9(35)$ which is favored by our data for 
$\chi_t$ obtained from lattice. The background orange band shows the corresponding bound within 
DIGA when the $K$-factor is varied between $1$-$2.5$ whereas the red band is estimated using the continuum 
result of $\chi_t$ taken from Ref. \cite{Bonati:2015vqz} considering a $1\sigma$ spread of the exponent $b$. 
As evident from Fig. \ref{fig:axiondm}, the $T_{osc}$ is extremely sensitive to the characteristic 
temperature exponent $b$ of the $\chi_t$ data. The values of $T_{osc}$ obtained from our analysis for any 
fixed value of $f_a$ within the range $10^{11}-10^{13}$ GeV, is atleast a factor five lower than the 
corresponding $T_{osc}$ extracted using data from Ref. \cite{Bonati:2015vqz}. The best estimate of the 
bound for $f_a$ from our analysis comes out to be $f_a\leq 1.2\times 10^{12}$ GeV with a ten percent error. 
The current uncertainties within the DIGA which is reflected in the magnitude of the $K$-factor would only 
mildly effect the estimates of the bound for $f_a$, changing it by $\sim 20\%$ when the $K$-factor varies between $1$-$2.5$ 
as seen in Fig. \ref{fig:axiondm}. This range of $f_a$ is to be probed from the measurements of axion mass in 
the current and next generation ADMX experiments \cite{ADMX}.

\section{Conclusions}
In summary, we have presented the continuum extrapolated results for topological susceptibility 
in 2+1 flavor QCD for $T\lesssim 3~T_c$ from first principles lattice calculations using 
Symanzik flow technique. The exponent $b$ that controls the fall-off of $\chi_t^{1/4}$ with 
temperature is found to be distinctly different in temperature intervals below and above $250$ MeV. 
For $T>250$ MeV, we show that $b\sim 2$, in agreement with the calculations when considering that the 
topological properties of QCD can be well described as a dilute gas of instantons. 
The main conclusion from this work is that the $\chi_t^{1/4}$ in QCD determined non-perturbatively 
for $T>250$ MeV can be very well described with the corresponding (partial) two-loop result within 
DIGA when the later is scaled up with a factor $K=1.90(35)$. This scaling factor is similar to the 
recent estimate in pure gauge theory \cite{Borsanyi:2015cka}. 

Our results are significantly lower in magnitude than the other continuum estimate of $\chi_t$ in QCD 
with dynamical fermions \cite{Bonati:2015vqz} for $T>250$ MeV.  This could possibly 
due to the fact that the previous study uses relatively coarser lattices. We show that the cut-off 
effects in $\chi_t$ depends on $a T$ rather than $a \Lambda_{QCD}$ and hence extremely sensitive to 
the lattice spacing. Considering this fact, our continuum extrapolation has been carefully 
performed. We also calculated the disconnected part of 
chiral susceptibility which gives an independent estimate for $\chi_t$ in the high temperature phase of 
QCD when the chiral symmetry is effectively restored. The continuum estimate of $\chi_t$ from this 
observable matches with our result using the gauge definition of topological charge which gives us 
further confidence in our procedure.

It would be desirable to check how sensitive is the scale factor $K$ on the matching procedure,
particularly when the matching of the lattice result for $\chi_t$ with the one within DIGA is 
performed at a temperature $\sim 1$ GeV. With the present day algorithms, we do not observe any 
signal for $\chi_t$ at finer lattice spacings for $T>450$ MeV hence it is necessary to 
explore new algorithms that sample different topological sectors more efficiently 
at higher temperatures (see e.g. \cite{Mages:2015scv}). However for axion phenomenology, the 
effects due to large uncertainties in $K$-factor is fairly mild. A change in the magnitude 
from $K=2.5$ to $K=1$, would reduce the estimates of axion decay constant by $20\%$ which is 
within the current experimental uncertainties for axion detection. With the temperature 
dependence of $\chi_t$ for $T>1.5~T_c$ now fairly well understood in terms of the microscopic 
topological constituents, it would be important to study in detail the topological properties 
in the vicinity of the chiral crossover transition in QCD.

\section*{Acknowledgements}
This work was supported by U.S. Department of Energy under Contract No. DE-SC0012704.
The calculations have been carried out on the clusters of USQCD collaboration.
We thank Frithjof Karsch, Swagato Mukherjee and Robert Pisarski for many helpful 
discussions. SS is grateful to  S\"oren Schlichting and Hooman Davoudiasl for very 
interesting discussions.

\bibliographystyle{elsarticle-num}
\bibliography{ref}

\end{document}